\documentclass[aps,prb,reprint,superscriptaddress]{revtex4-1}

\usepackage{graphicx}

\begin{document}

\title{Switching of magnetic domains reveals evidence for spatially inhomogeneous superconductivity}

\author{Simon~Gerber}
\affiliation{Laboratory for Neutron Scattering, Paul Scherrer Institut, CH-5232 Villigen, Switzerland}
\author{Marek~Bartkowiak}
\affiliation{Laboratory for Developments and Methods, Paul Scherrer Institut, CH-5232 Villigen, Switzerland}
\author{Jorge~L.~Gavilano}
\affiliation{Laboratory for Neutron Scattering, Paul Scherrer Institut, CH-5232 Villigen, Switzerland}
\author{Eric~Ressouche}
\affiliation{SPSMS, UMR-E CEA/UJF-Grenoble 1, INAC, F-38054 Grenoble, France}
\author{Nikola~Egetenmeyer}
\author{Christof~Niedermayer}
\affiliation{Laboratory for Neutron Scattering, Paul Scherrer Institut, CH-5232 Villigen, Switzerland}
\author{Andrea~D.~Bianchi}
\affiliation{D\'{e}partement de Physique \& RQMP, Universit\'{e} de Montr\'{e}al, Montr\'{e}al, Qu\'{e}bec H3C 3J7, Canada}
\author{Roman~Movshovich}
\author{Eric~D.~Bauer}
\author{Joe~D.~Thompson}
\affiliation{Condensed Matter and Magnet Science, Los Alamos National Laboratory, Los Alamos, New Mexico 87545, USA}
\author{Michel~Kenzelmann}
\affiliation{Laboratory for Developments and Methods, Paul Scherrer Institut, CH-5232 Villigen, Switzerland}

\begin{abstract}
\begin{center}\noindent Publication reference: Nature Physics \textbf{10}, 126--129 (2014); DOI:10.1038/nphys2833 \\www.nature.com/nphys/journal/v10/n2/abs/nphys2833.html
\end{center}
\end{abstract}

\maketitle

\textbf{The interplay of magnetic and charge fluctuations can lead to quantum phases with exceptional electronic properties. A case in point is magnetically-driven superconductivity\cite{mathur98,monthoux07}, where magnetic correlations fundamentally affect the underlying symmetry and generate new physical properties. The superconducting wave-function in most known magnetic superconductors does not break translational symmetry. However, it has been predicted that modulated triplet \mbox{\textit{p}-wave} superconductivity occurs in singlet \textit{d}-wave superconductors with spin-density wave (SDW) order\cite{aperis10,agterberg09}. Here we report evidence for the presence of a spatially inhomogeneous \textit{p}-wave Cooper pair-density wave (PDW) in CeCoIn$_5$. We show that the SDW domains can be switched completely by a tiny change of the
magnetic field direction, which is naturally explained by the presence of triplet superconductivity. Further, the \textit{Q}-phase emerges in a common magneto-superconducting quantum critical point. The \mbox{\textit{Q}-phase} of CeCoIn$_5$ thus represents an example where spatially modulated superconductivity is associated with SDW order.}

Superconductivity emerges in solids when electrons form Cooper pairs at low temperatures and condense in a macroscopically coherent quantum state, leading to zero-resistance and diamagnetism. For most phonon-mediated superconductors, the symmetry of the superconducting gap function is an isotropic singlet \mbox{\textit{s}-wave} that does not break the symmetries of the underlying lattice. Unconventional superconductors\cite{norman11,sigrist91} adopt more complex wave-functions: singlet \textit{d}-wave in cuprates, extended \textit{s}-wave for the pnictides and \textit{p}-wave in ruthenates\cite{mackenzie03}. Although co-existing wave-functions are theoretically possible, most superconductors are adequately described by just a single wave-function. Here we provide evidence that \textit{p}-wave superconductivity emerges in a \textit{d}-wave superconductor and couples to a SDW---forming a unique quantum phase with a spatially inhomogeneous superconducting Cooper pair-density wave.

The stoichiometric heavy-fermion superconductor CeCoIn$_5$ serves as a model material for studies of \mbox{$d_{x^2-y^2}$-wave} superconductivity\cite{petrovic01,thompson12}. It shares many properties with the cuprate superconductors, such as the zero-field superconducting wave-function\cite{movshovich01,izawa01,curro01} symmetry, a quasi-two-dimensional Fermi surface\cite{settai01} and a spin resonance in the superconducting state\cite{stock08}. Superconductivity emerges below $T_{\rm c}(0)$~=~2.3~K and is Pauli-limited. The superconducting phase transition becomes first-order\cite{bianchi03} close to $H_{\rm c2}$(0), where SDW order has been observed that only exists in the \textit{Q}-phase situated inside the superconducting phase\cite{young07,kenzelmann08,kenzelmann10}. The SDW propagation is pinned along the $d_{x^2 - y^2}$-line nodes, where low-energetic quasiparticles allow for electron nesting\cite{kenzelmann10} (see Fig.~1a). 

\begin{figure*}
	\includegraphics[width=0.743\linewidth]{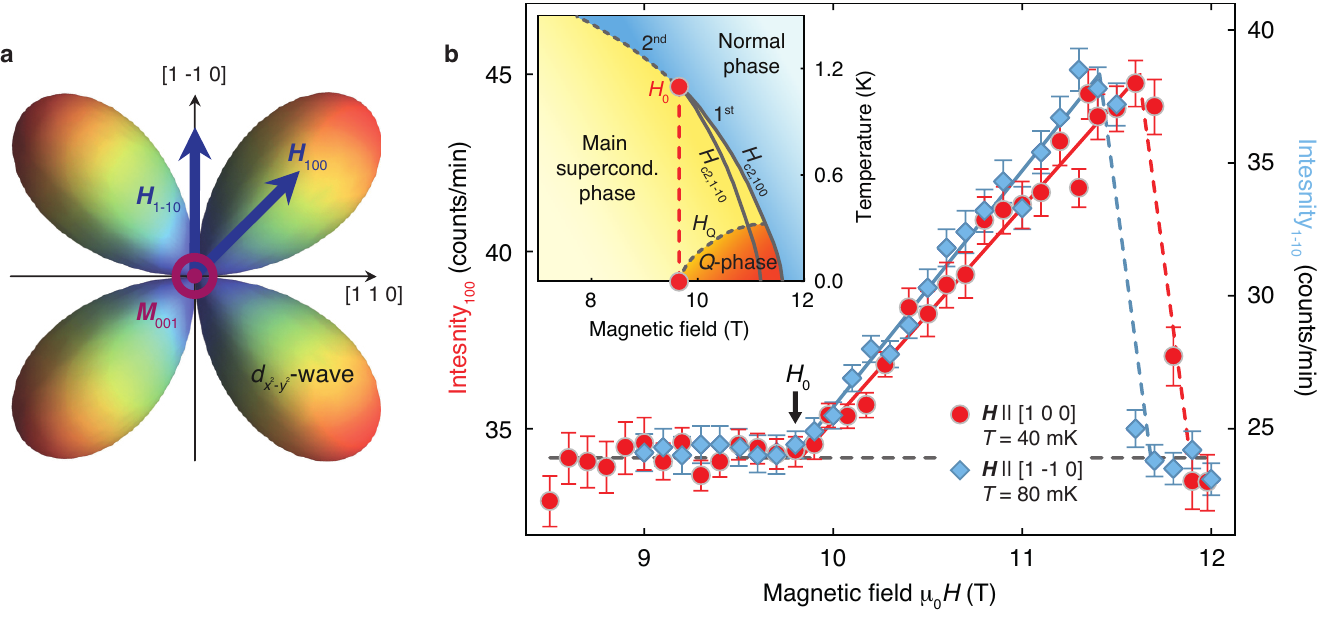}
	\caption{\textbf{A novel magneto-superconducting quantum critical point. a}, $d_{x^2 - y^2}$-superconducting wave-function and external magnetic field \textbf{\textit{H}} along the nodal [1~-1~0] and fully gapped [1~0~0] direction, where identical SDW order was detected\cite{kenzelmann08,kenzelmann10}. \textbf{b}, The nature of the lower \textit{Q}-phase boundary for \textbf{\textit{H}}~$||$~[1~-1~0] (blue diamonds) and [1~0~0] (red circles) was investigated by probing the magnetic Bragg peak intensity as a function of $H$. Magnetism appears in a continuous second-order phase transition. Thus, a magnetically driven quantum critical point is located inside the superconducting phase at $H_Q$(0). The onset field $\mu_0H_Q\approx$~9.8~T, coincides with the field strength $H_0$, where the superconducting transition becomes first-order and anisotropic (see inset). Solid blue and red lines depict linear fits to the intensity \textit{I}$_{\rm 1-10}$ and \textit{I}$_{\rm 100}$, respectively. Blue, red and grey dashed lines are guides to the eye for \textit{H}$_{\rm c2, 1-10}$, \textit{H}$_{\rm c2,100}$ and the background, respectively. Error bars, 1 standard deviation~(s.d.).}
\end{figure*}

In some microscopic theories addressing the nature of the \textit{Q}-phase, the formation of a spatially modulated Fulde--Ferrell--Larkin--Ovchinnikov (FFLO) state plays a central role\cite{yanase11}. Other theoretical approaches emphasize SDW nesting\cite{suzuki11,kato11}. Phenomenologically it has been shown\cite{kenzelmann08,aperis10} that SDW order in a \textit{d}-wave superconductor always couples to a PDW of mixed singlet/triplet nature. The precise symmetry properties of possible coupling terms have been presented by Agterberg \textit{et al}.\cite{agterberg09}:
\begin{equation}
	V_1\propto i M_{Q}^z (\Delta_0^{\star} \Delta_{-Q} - \Delta_0 \Delta_{Q}^{\star})~+~{\rm c.c.}\label{Eq1}
\end{equation}
\begin{equation}
	V_2\propto H M_{Q}^z (\Delta_0^{\star} \Delta_{-Q} + \Delta_0 \Delta_{Q}^{\star})~+~{\rm c.c.},\label{Eq2}
\end{equation}
where $\Delta_0$, $M_{Q}^z$ and $\Delta_{-Q}$ represent the $d_{x^2-y^2}$, SDW and PDW order parameter, respectively. The finite-momentum ($-Q$) triplet PDW component is required to conserve momentum (translational symmetry). An FFLO state would be only weakly modulated, \mbox{$q_{\rm FFLO}<Q$}, and so a phenomenological interaction coupling to FFLO order in the presence of an SDW would require additional, higher-order coupling terms.

We searched for evidence of coupling terms $V_1$ or $V_2$ through an investigation of the lower \textit{Q}-phase boundary. Figure~1b shows a high-statistics measurement of the \textit{H}-dependence of the magnetic Bragg peak intensity for \textbf{\textit{H}}~$||$~[1~0~0] and [1~-1~0]. The onset of the magnetic scattering occurs at a continuous phase transition at $\mu_0H_Q(0)\approx$~9.8~T, is independent of the field direction in the basal plane and grows linearly with field. It corresponds also to the field strength $H_0\approx H_Q(0)$, where the superconducting transition becomes first-order\cite{bianchi03}. The magnetic neutron intensity is proportional to the square of the magnetic moment $I\propto M^2$, i.e., the measured SDW order parameter grows as $M\propto (H/H_0-1)^\beta$ with a critical exponent $\beta \approx0.5$---as expected for a quantum phase transition. Our data excludes an additional phase\cite{koutroulakis10} adjacent to the \textit{Q}-phase.

Microscopic theories that rely on nesting and an increased density of states naturally lead to the coexistence of \textit{Q}-domains or the emergence of so-called double-\textit{Q}-phases, particularly for \textit{\textbf{H}}~$||$~[1~0~0]. The two possible SDW domains \textit{Q}$_{\rm h}$ and \textit{Q}$_{\rm v}$, associated with \textbf{\textit{Q}}~=~($q$,~$\pm q$,~0.5), are shown in figure~2a-c. For fields \textit{\textbf{H}}~$||$~[1~-1~0], we find all four horizontal Bragg peaks associated with \textit{Q}$_{\rm h}$, one of which is shown in figure~2d. However, no peaks belonging to the \textit{Q}$_{\rm v}$-domain, as for example (0.44,~--0.44,~0.50), were detected (see Fig.~2e), showing that only one of the two SDW domains is populated for fields along the nodal direction. Even for a field along the anti-nodal direction (\textbf{\textit{H}}~$||$ ~[1~0~0]), where both \textit{Q}-domains are identical by symmetry, we find a mono-domain state (see supplementary discussion). The angle~$\psi$ between \textbf{\textit{H}} and [1~0~0] was smaller than $0.2^\circ$. This is evidence against the available microscopic theories and suggests the presence of an additional order parameter such as a PDW.

\begin{figure*}
	\includegraphics[width=\linewidth]{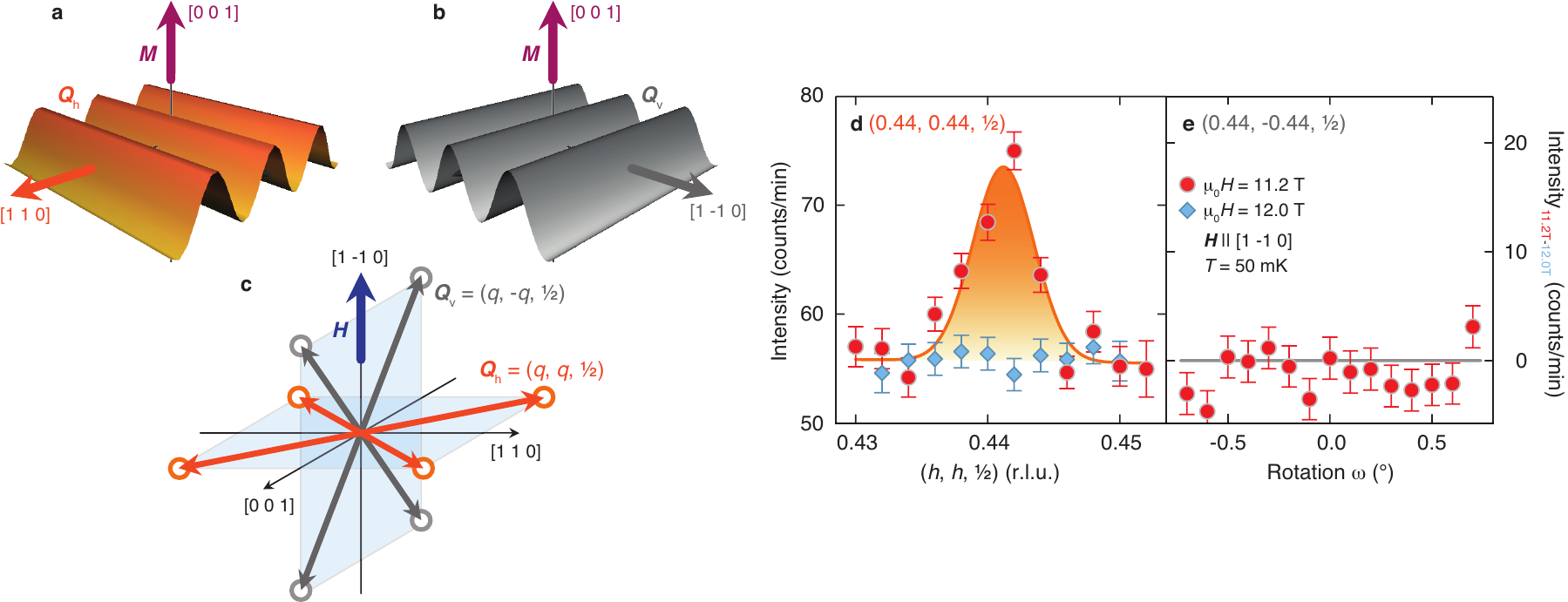}
	\caption{\textbf{Imbalance of the \textit{Q}-phase domain population. a}, \textbf{b}, SDW domains in real space: Magnetic moments \textbf{\textit{M}}~$||$~[0~0~1] are modulated perpendicular to the orthogonal propagation vectors \textbf{\textit{Q}}$_{\rm h}$ and \textbf{\textit{Q}}$_{\rm v}$. \textbf{c}, Corresponding magnetic Bragg positions form an eightfold star of \textit{Q}-vectors in reciprocal space. The orange subset of equivalent positions \textbf{\textit{Q}}$_{\rm h}$~=~($q$,~$q$,~0.5) lies completely in the horizontal scattering plane, whereas the grey \textbf{\textit{Q}}$_{\rm v}$~=~($q$,~--$q$,~0.5) domain has vertical components \textit{\textbf{H}}~$||$~[1~-1~0]. \textbf{d}, Diffracted intensity of the $Q_{\rm h}$-domain at $T$~=~50~mK and $\mu_0H$~=~11.2~T (red circles). Magnetic scattering is absent in the normal state ($\mu_0H$~=~12.0~T, blue diamonds). The scan is shown in reciprocal lattice units (r.l.u.). \textbf{e}, Background-subtracted diffracted intensity $I_{\rm 11.2~T-12.0~\rm T}$ (red circles) of a rotation $\omega$-scan at (0.44,~-0.44,~0.5), with a moving average of nearest neighbours for the normal state data, indicates an unpopulated $Q_{\rm v}$-domain. Lines depict Gaussian fits to the data. Error bars, 1 s.d.}
\end{figure*}

The observation of only one SDW domain for \mbox{\textbf{\textit{H}}~$||$ ~[1~0~0]} points towards a hypersensitivity of the \mbox{\textit{Q}-phase} with respect to \textbf{\textit{H}}. To determine the switching field direction, we equipped the dilution refrigerator with a piezoelectric \textit{attocube} sample rotator. Figure~3a shows the integrated Bragg peak intensity of the \textit{Q}$_{\rm h}$- and \textit{Q}$_{\rm v}$-domains as a function of the angle $\psi$ (see Fig.~3b). The domain population can be fully switched within only $\Delta\psi\approx 0.1^\circ$.

\begin{figure*}
	\includegraphics[width=\linewidth]{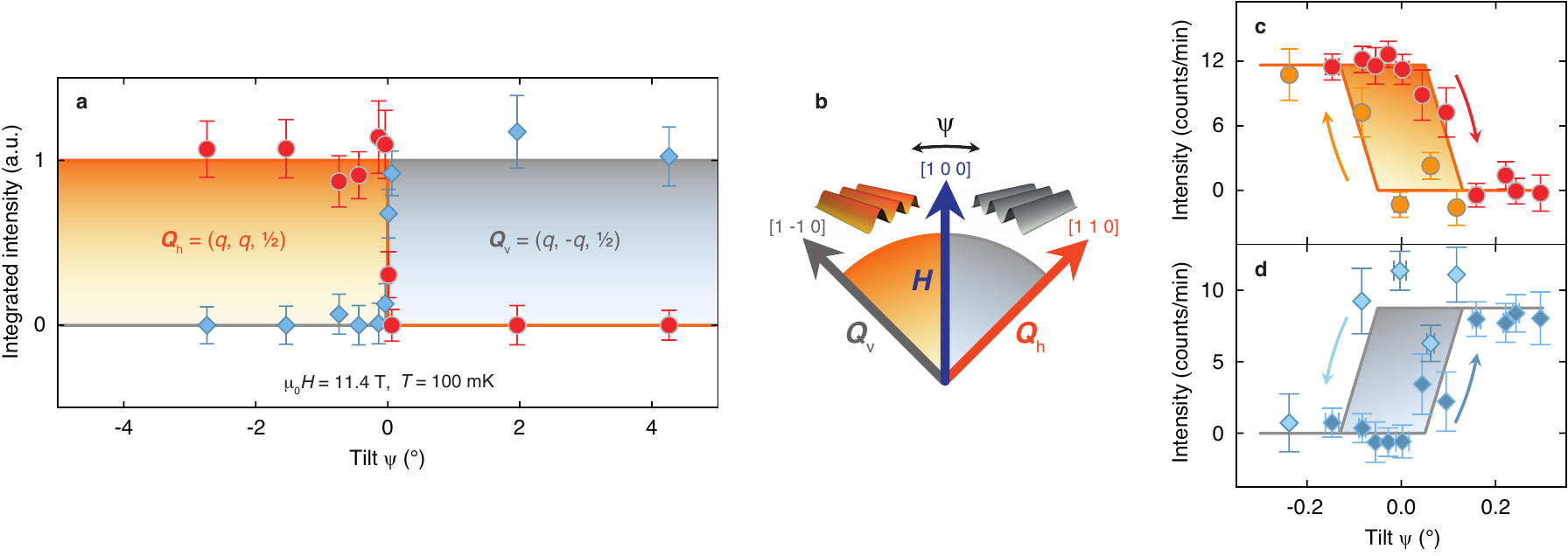}
	\caption{\textbf{Switching of magneto-superconducting domains. a}, Integrated intensity of the $Q_{\rm h}$- (red circles) and the \mbox{$Q_{\rm v}$-domain} (blue diamonds), normalized to the averaged respective full population, as a function of the tilt angle $\psi$ (see sketch~\textbf{b}). A mono-domain population is observed for $|\psi| \geq0.05^\circ$, which can be macroscopically switched in a sharp transition. Thus, the magnetic field direction allows direct access and control of the complex quantum state in the \textit{Q}-phase. The magneto-superconducting state was newly prepared for each field direction. \textbf{c}, \textbf{d}, Measurement of the hysteresis $\psi\approx0.2^\circ$ of the first-order domain switch by field rotation without leaving the \textit{Q}-phase. Data measuring while increasing (decreasing) the tilt angle are shown in red (orange) circles and dark (light) blue diamonds. Lines and shaded areas are guides to the eyes. Error bars, 1 s.d.}
\end{figure*}

The nature of the SDW switch was further probed at ($q$,~$\pm q$,~0.5) by rotating \textbf{\textit{H}} without leaving the \textit{Q}-phase. We find a first-order domain switching with a hysteresis of only $\Delta\psi\approx0.2^\circ$ (see Fig.~3c,d). This provides evidence that the two coexistent single-\textit{Q} SDW domains are not in a coherent superposition (double-\textit{Q}-state) for \mbox{$|\psi| \leq0.05^\circ$}, since the domain population changes hysteretically from unpopulated to populated.

\begin{figure*}
	\includegraphics[width=\linewidth]{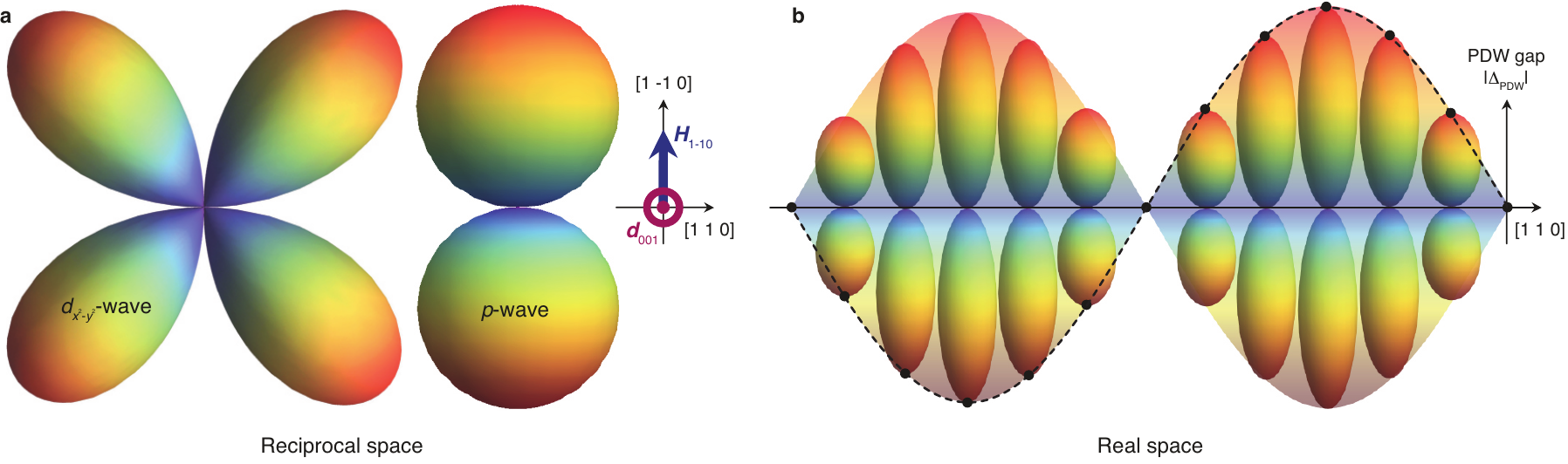}
	\caption{\textbf{A Cooper pair-density wave in the \textit{Q}-phase. a}, A phenomenological lowest-order linear coupling term of $d_{x^2-y^2}$-superconductivity and incommensurate SDW order requires, as a result of momentum conservation, the appearance of an additional triplet PDW component. The constraints in the \textit{Q}-phase allow only one particular triplet state, represented by a linear combination of two dumbbell-shaped $p$-wave gap functions. The schematic shows the $d_1$-triplet component for \textit{\textbf{H}}~$||$~[1~-1~0], where electron nesting of low-energetic quasiparticles is only possible along the nodes perpendicular to the field direction (\textit{\textbf{Q}}~$\perp$~\textit{\textbf{H}}). The triplet component also accounts for the hypersensitive switching of the SDW close to \textit{\textbf{H}}~$||$~[1~0~0]. Shown is the point symmetry of the gap functions. The nodes of both the \textit{d}- and \textit{p}-wave superconducting gap functions are in reciprocal space. \textbf{b}, Representation of the local PDW gap structure along the [1~1~0] direction in real space. The Cooper pair-density wave is modulated in real space as a result of the direct coupling to the incommensurate SDW order. This produces additional nodes in real space where the amplitude of the $p$-wave gap function is zero.}
\end{figure*}

The observed hypersensitivity of the \textit{Q}-domain on the magnetic field direction is remarkable. A Zeeman term \textit{\textbf{M}}$\cdot$\textit{\textbf{H}} that controls magnetic domains in other SDW materials, such as Cr or MnSi, can be excluded as the origin: the SDW ordered moments \textit{\textbf{M}}~$||$~[0~0~1] are perpendicular to \textit{\textbf{H}} for all fields in the tetragonal plane\cite{kenzelmann08,kenzelmann10}. Alternatively, if the \textit{Q}-phase were FFLO driven, it is possible that weakly localized magnetic moments at FFLO nodal planes\cite{yanase11} could lead to SDW mono-domains through the difference of the horizontal and vertical SDW coherence length. However, this scenario does not easily explain the hypersensitivity of the mono-domain state around [1~0~0], which would require a fine-tuning of interactions. Similar modifications would be necessary if the \textit{Q}-phase were purely nesting driven\cite{suzuki11}.

A simpler scenario, where a SDW and PDW are coupled in the presence of the \textit{d}-wave condensate is described by equation~(\ref{Eq1}), which involves singlet/triplet mixing. The symmetry of the PDW is highly constrained: for \textbf{\textit{H}}~$||$~[1~-1~0] and \textbf{\textit{Q}}$_{\rm h}$~=~($q$,~$q$,~0.5) the allowed \textit{p}-wave component is a linear combination of two spin-triplet pairing vector functions \textbf{\textit{d}}$_1$(\textbf{\textit{k}})~=~(0,~0,~$k_x$--$k_y$) (nodes along [1~1~0], see Fig.~4a) and \textbf{\textit{d}}$_2$(\textbf{\textit{k}})~=~($k_z$,~--$k_z$,~0) (nodal plane ($h$,~$l$,~0)). Since the direction of the SDW \textit{Q}-vector and the nodal direction of the \textit{p}-wave-function are identical, the PDW and SDW can naturally coexist and couple. In contrast, the gap functions\cite{agterberg09} allowed by the alternate coupling in equation~(\ref{Eq2}) do not produce nodes along [1~1~0], and thus compete with SDW formation.

Experimentally, we observe that electron nesting in the \textit{Q}-phase occurs always along the $d_{x^2-y^2}$-node direction that is more perpendicular to the applied field (\textit{\textbf{Q}}~$\perp$~\textit{\textbf{H}}). Triplet components can feature a strongly anisotropic magnetic susceptibility, as in the \textit{A}-phase\cite{leggett75} of $^3$He and also for \textbf{\textit{d}}$_2$(\textbf{\textit{k}}). This provides a simple microscopic mechanism for the switching behaviour: \textit{\textbf{H}}~controls the \mbox{\textit{p}-wave} line node, which simultaneously determines the direction of the SDW wave vector. Since some components of the magnetic susceptibility tensor of \mbox{\textit{p}-wave} superconductivity do not change compared to the normal phase, there are nuclear magnetic resonance lines (NMR) at the position of the normal state\cite{kumagai11}. Our arguments are bolstered by a significantly enhanced spin susceptibility\cite{mitrovic06} in the \textit{Q}-phase, compared to the main \textit{d}-wave phase, and by measurements of the quasiparticle entropy\cite{tokiwa12}. The switching behaviour of the SDW domains should also affect the thermal conductivity, making it anisotropic and hypersensitive to the field direction.

NMR identified low-energy quasiparticles\cite{kumagai11} whose density of states scales with a critical exponent $\beta \approx0.5$, identical to that of the SDW that we observe in our present study. This suggests that the transition at $H_Q(0)$ also affects the electronic structure, and may constitute a magneto-superconducting quantum critical point involving both superconducting and magnetic degrees of freedom. A linear coupling term as described in equation~(\ref{Eq1}) may thus indeed be present.

Our results suggest that the quantum critical point for fields in the tetragonal plane\cite{ronning05} near $H_{\rm c2}(0)$ is located at $H_Q(0)$ and involves both magnetic and charge degrees of freedom. It is characterized with a relatively subtle change of the electronic structure compared to what is expected in a Kondo breakdown scenario. Probably a similar scenario can also explain the quantum critical point\cite{bianchi03ii,paglione03} observed for fields along the tetragonal axis, where an anisotropic violation of the Wiedemann--Franz law points towards an anisotropic destruction of the Fermi surface\cite{tanatar07}.

The sharp switching of magneto-superconducting domains provides evidence for the condensation of a spatially modulated triplet PDW (see Fig.~4b), which emerges spontaneously with a SDW at $H_Q(T)$. Our experiment does not allow a distinction of whether the SDW or the \textit{p}-wave PDW drives the \textit{Q}-phase. A non-unitary extension of the proposed \textit{d}-vector representatives\cite{agterberg09} may suggest a related scenario, where SDW moments represent the ordering of ferromagnetic components of the Cooper pair-density wave.

Our experiment confirms the general prediction that an incommensurate SDW in a singlet superconductor can couple to a spatially inhomogeneous triplet PDW\cite{aperis10}. The method showcases a new approach to study the properties of the quantum condensate in a superconductor. Finally, the observed switching of magneto-superconducting domains and the manipulation of a quantum state using magnetic fields may represent a promising approach for the encoding of quantum information in the solid state.\\

\small
\noindent\textbf{Methods summary}\\High-field neutron diffraction experiments were carried out on the thermal neutron two-axis diffractometer D23 at the Institut Laue-Langevin, Grenoble, France and the cold neutron triple-axis spectrometer RITA-II at the Swiss Spallation Neutron Source SINQ, Paul Scherrer Institut, Villigen, Switzerland. The same CeCoIn$_5$ single-crystal with a mass of 155~mg was used on both instruments. Temperatures down to $T$~=~40~mK and external magnetic fields of up to $\mu_0H$~=~12.0~T were reached with vertical-field cryomagnets and dilution refrigerator inserts. D23 is equipped with a lifting-arm detector, which allows measuring a wave vector transfer with a vertical component, i.e., along the magnetic-field direction. Measurements on D23 were carried out with an incident neutron wavelength of $\lambda_{\rm n}$~=~1.28~\AA~obtained from the (2,~0,~0) Bragg reflection of a flat Cu monochromator. $\lambda_{\rm n}$~=~4.04~\AA~from the (0,~0,~2) reflection of a vertically focusing pyrolitic graphite (PG) monochromator was used on RITA-II. There, a beryllium filter in front of the nine-bladed PG multi-analyser minimized contributions of higher order neutrons. Data shown in figure~1 were recorded by keeping the diffractometer at the Bragg position, whilst ramping~$H$. This is justified, since the peak position moves much less than its resolution. \textit{q}-scans are depicted in ($h$,~$k$,~$l$) reciprocal lattice units (r.l.u.). Measurements with a rotating field direction were carried out on the instrument D23 by utilizing a non-magnetic piezoelectric sample rotator (type \textit{ANGt50} from \textit{attocube systems AG}) inside the dilution refrigerator. The field direction was measured by a built-on Hall probe as well as via the vertical tilt of (0,~$k$,~$l$) nuclear Bragg peaks. An additional thermometer located next to the sample allowed for an accurate measurement of the sample temperature (see supplementary methods).\\

\noindent\textbf{Additional information}\\Correspondence and requests for materials should be addressed to michel.kenzelmann@psi.ch.\\

\noindent\textbf{Acknowledgements}\\This work is based on neutron scattering experiments performed at the Institut Laue-Langevin, Grenoble, France and the Swiss Spallation Neutron Source SINQ, Paul Scherrer Institut, Villigen, Switzerland. We thank P.~Fouilloux and M.~Zolliker for technical assistance. Discussions with M.~Sigrist as well as C.~Batista, P.~Coleman, K.~Machida, K.~Kumagai, and J.~S.~White are acknowledged. This work was supported by the Swiss NSF (Contract No. \mbox{200021-122054}, 200020-140345 and MaNEP). A.D.B. received support from NSERC, FQRNT and the Canada Research Chair Foundation. Work at LANL was performed under the auspices of the US DOE, Office of Basic Energy Sciences, Division of Materials Sciences and Engineering.\\

\noindent\textbf{Author contributions}\\S.G. and M.K. conceived and led the project. S.G., M.B., J.L.G., E.R., N.E., C.N., and M.K. carried out the experiments. M.B. incorporated the piezoelectric sample rotator into the setup. E.D.B and J.D.T. grew and characterized the CeCoIn$_5$ single-crystal. S.G. analysed the data. S.G., J.L.G and M.K. wrote the manuscript with input from all co-authors.\\

\noindent\textbf{Competing financial interests}\\The authors declare no competing financial interests.\\

\widetext
\clearpage
\normalsize
\begin{center}
	{\large\textit{Supplementary information}\\ \textbf{Switching of magnetic domains reveals evidence for spatially inhomogeneous superconductivity\\}}
	\vspace{4mm}\noindent Simon~Gerber, Marek~Bartkowiak, Jorge~L.~Gavilano, Eric~Ressouche, Nikola~Egetenmeyer, Christof~Niedermayer, Andrea~D.~Bianchi, Roman~Movshovich, Eric~D.~Bauer, Joe~D.~Thompson, and Michel~Kenzelmann\\
	\vspace{2.5mm}Publication reference: Nature Physics \textbf{10}, 126--129 (2014); DOI:10.1038/nphys2833 \\www.nature.com/nphys/journal/v10/n2/abs/nphys2833.html\end{center}
\vspace{3mm}
\twocolumngrid

\subsection*{\normalsize\normalfont \textit{Supplementary discussion}\\ \textbf{Spin-density wave domains in the \textit{Q}-phase}}

We briefly explain the \textit{Q}-phase state in terms of SDW domains. A real space representation of the two possible SDW structures is shown in figure~2a and 2b. Both the \textit{Q}$_{\rm h}$- (orange) and the \textit{Q}$_{\rm v}$-domain (grey) have modulated magnetic moments \textbf{\textit{M}}~$||$~[0~0~1], but propagate in orthogonal directions (arrows). In reciprocal space, the two domains are described by two subsets of \textit{Q}-vectors forming an eightfold star of \textit{Q}-vectors as schematically depicted in figure~2c of the main article. By mirror and translational symmetry, the eight \textit{Q}-vectors can be attributed to either the horizontal \textbf{\textit{Q}}$_{\rm h}$~=~($q$,~$q$,~0.5) or the vertical \textbf{\textit{Q}}$_{\rm v}$~=~($q$,~--$q$,~0.5) wave vector as shown in orange and grey, respectively. However, the two domains are not equivalent in the depicted geometry with \textbf{\textit{H}}~$||$~[1~-1~0]: \textbf{\textit{Q}}$_{\rm v}$ has components parallel to \textbf{\textit{H}} and \textbf{\textit{Q}}$_{\rm h}$ only perpendicular components. For \textbf{\textit{H}}~$||$~[1~0~0], the two domains are degenerate with equal components parallel to the field direction (see sketch in Fig.~S\ref{singleq100}b).

A \textit{q}-scan along [1~-1~0] around ($0.44$,~$0.44$,~0.5) at $\mu_0H$~=~11.2~T confirms long-range magnetic order along \textit{\textbf{H}}~$||$~[1~-1~0], with a lower bound of the vertical correlation length of 300~\AA. Three-dimensional long-range SDW order was found earlier\cite{kenzelmann10ii} also for \textit{\textbf{H}}~$||$~[1~0~0].

\subsection*{\normalsize\normalfont \textit{Supplementary discussion}\\\textbf{Single-\textit{Q} spin-density wave order for \textbf{\textit{H}}~$||$~[1~0~0]}}

\begin{figure}
	\centering
	\includegraphics[width=\linewidth]{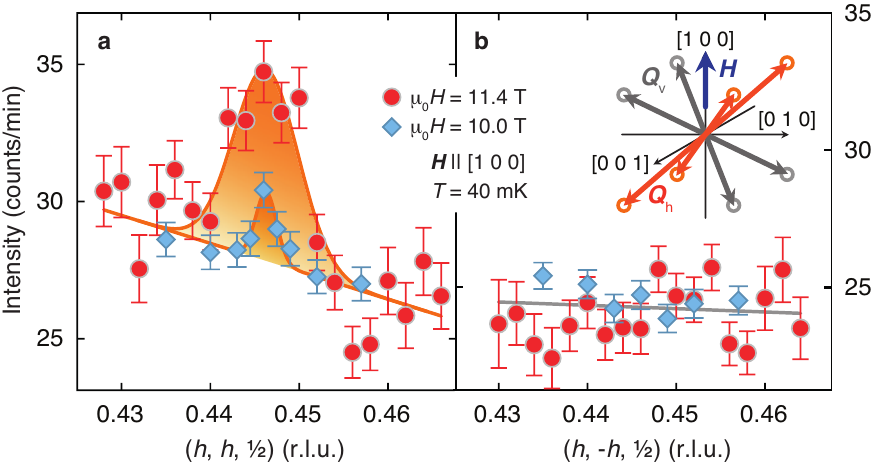}
	\caption{\label{singleq100}
		 (supplementary) \textbf{Mono-domain single-\textit{Q} magnetism for \textit{H}~$||$~[1~0~0].} The two spin-density wave domains are degenerate with respect to the field direction, the crystal structure and the $d_{x^2-y^2}$-wave gap function when \textbf{\textit{H}}~$||$~[1~0~0] (see sketch in \textbf{b}). Within the accuracy of the alignment, a mono-domain population is found as well: magnetic Bragg peaks appear only at $Q_{\rm h}$- (\textbf{a}) but not at $Q_{\rm v}$-positions (\textbf{b}). Single-\textit{Q}, long-range ordered magnetism is also observed close to the lower \textit{Q}-phase boundary ($\mu_0H$~=~10.0~T, blue diamonds). Scans are shown in reciprocal lattice units (r.l.u.). Lines depict Gaussian fits to the data with a common background for the $\mu_0H$~=~10.0 and 11.4~T data. Error bars,~1~s.d.}
\end{figure}

Only one SDW domain was examined for the determination of the magnetic structure of the \textit{Q}-phase in the earlier study\cite{kenzelmann10ii} with \textbf{\textit{H}}~$||$~[1~0~0]. Since this field direction does not break the symmetry equivalence of the two magnetic domains (see sketch in Fig.~S\ref{singleq100}b), it was expected that both magnetic domains are present. On the diffractometer D23 we revisited this case: 18 nuclear reflections were used in the alignment and a precision of 0.06$^\circ$ and 0.15$^\circ$ with respect to the magnetic field direction was obtained in the direction of [0~1~0] and [0~0~1], respectively. It is important to note that only a misalignment along [0~1~0] affects the degeneracy of the SDW domains as a result of the commensurability along [0~0~1]. We probed the domain population in this geometry by searching for magnetic scattering at the top four positions of the eightfold star of \textit{Q}-vectors.

Data presented in figure~S\ref{singleq100} show the surprising experimental result, that for \textbf{\textit{H}}~$||$~[1~0~0] only the $Q_{\rm h}$-domain is populated, but not the $Q_{\rm v}$-domain. Preparing the magnetic state at $\mu_0H$~=~11.4~T and $T$~=~50~mK by ramping $H$ from below the lower \textit{Q}-phase boundary or by quenching through the first-order superconducting transition did not result in a different domain population. In view of the precise sample alignment this points to a hypersensitivity of the domain population with respect to the magnetic field direction.

We take advantage of the mono-domain population, described by a single \textit{Q}-vector, to test whether a \mbox{double-\textit{Q}} state is realized at the lower \textit{Q}-phase boundary as theoretically suggested\cite{kato11ii,kato12}. A double-\textit{Q} structure is described by a coherent superposition of a $Q_{\rm v}$- and \mbox{$Q_{\rm h}$-modulation}, and thus magnetic scattering at both kinds of magnetic Bragg positions should be observed in the experiment. Double-\textit{Q} antiferromagnetism could, in principle, also account for the NMR line broadening\cite{koutroulakis10ii} observed in this field range. Data, represented by blue diamonds in figure~S\ref{singleq100}, were measured at $\mu_0H$~=~10.0~T, i.e., in the field range, where NMR data may be explained by a double-\textit{Q} structure. Our data shows a single-\textit{Q} SDW mono-domain at this field strength as well, which excludes the presence of an additional double-\textit{Q}-phase in the examined field range.

\subsection*{\normalsize\normalfont \textit{Supplementary methods}\\\textbf{Piezoelectric \textit{attocube} sample rotator}}

\begin{figure}
	\centering
	\includegraphics[width=\linewidth]{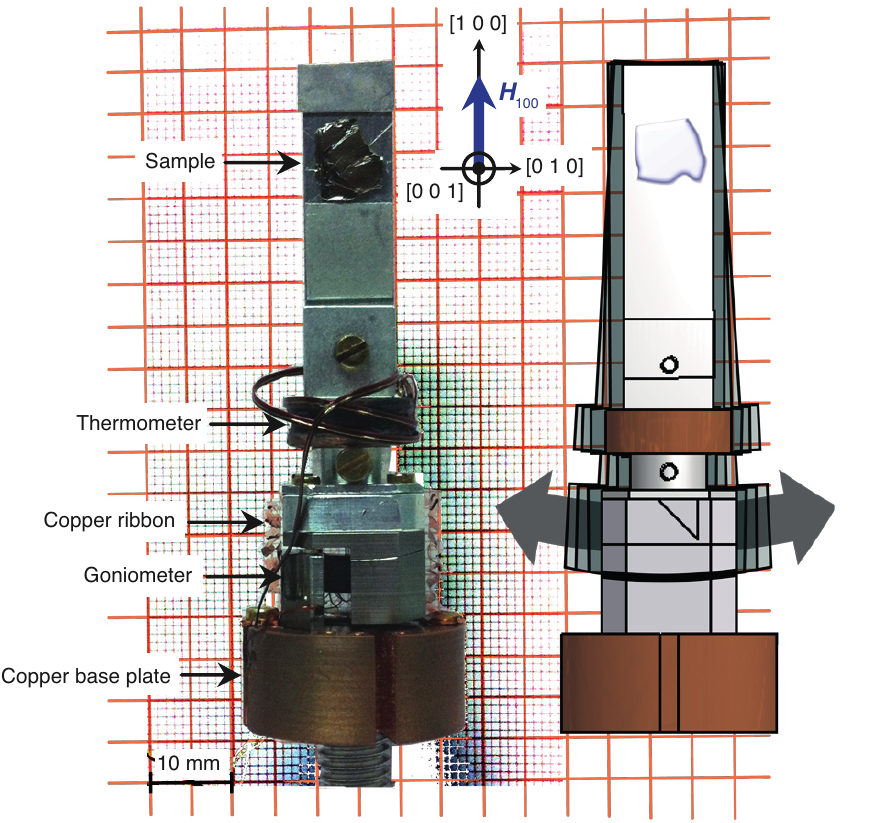}
	\caption{\label{gonio}
		 (supplementary) \textbf{Piezoelectric \textit{attocube} sample rotator.} The goniometer is fixed to a copper base plate and thermally linked to the mixing chamber of the dilution refrigerator. A sample platform is mounted on the moving part of the goniometer, which consists of an aluminium base plate for the ruthenium oxide thermometer, a Hall sensor and the sample holder. The single-crystalline CeCoIn$_5$ sample is glued to a 0.5~mm thick pure aluminium plate. Thermal anchoring of the sample assembly to the copper base plate is achieved by two corrugated copper ribbons.}
\end{figure}

Measurements described in the previous section imply that the switching between the two possible SDW domains happens in a very narrow range around \textbf{\textit{H}}~$||$~[1~0~0]. To reach the alignment precision needed, we devised a sample rotator which operates at low temperatures and high magnetic fields. Our setup is based on a purpose-built non-magnetic piezoelectric goniometer (type \textit{ANGt50} from \textit{attocube systems AG}), providing an angular range $\Delta\psi$~=~$\pm$3.6$^\circ$ and requiring a sample space diameter of only 20~mm---small enough to fit the 31~mm diameter available inside the \textit{Kelvinox-VT} dilution refrigerator. A copper base plate was used to fix the goniometer. The pre-aligned sample was glued to a 0.5~mm thick pure aluminium plate and mounted at the top of the goniometer together with a ruthenium oxide thermometer (see Fig.~S\ref{gonio}). The goniometer body is made from titanium and the rotor as well as the stator are mechanically only coupled via the piezo-crystal. Two corrugated copper ribbons were spanned between the base plate and the rotor of the goniometer to provide proper thermal contact between the sample and the dilution refrigerator. The ribbon was made from a 12~$\mu$m thick, 6~mm wide and 22~mm long pure copper foil (99.99+). By choosing a foil for the coupling rather than a wire, the cross section corresponds to a wire of 420~$\mu$m diameter, we ensure both good mechanical flexibility and good thermal conductivity. The whole setup was preassembled and the base plate was fixed to the copper rod extending from the mixing chamber of the dilution refrigerator towards the centre of the magnet, i.e., the neutron beam.

A base temperature of $T_{\rm s}$~=~$T_{\rm mix}$~=~38~mK was reached at zero magnetic field, whilst we obtained a sample temperature of $T_{\rm s}$~=~65~mK at $\mu_0 H$~=~11.4~T ($T_{\rm mix}$~=~40~mK at the mixing chamber). The elevated sample temperature in high magnetic fields can be explained by the additional heat input due to eddy currents and the reduction of the thermal conductivity of copper. Moreover, the increased specific heat of the assembly at high magnetic fields raised the equilibration times from about 10~minutes to over 1~hour.

The most dominant source of heat is the goniometer drive. Any movement of the goniometer introduces heat into the system, because a piezo-motor works on the principle of slip-stick friction. We operated the piezo-drive with an input voltage of 30~V and a repetition rate of 100~Hz. A full sweep took about 10~minutes and heated the sample to above $T_{\rm s}$~=~6~K while the mixing chamber reached $T_{\rm mix}$~=~1.2~K. Typical moving angles in the experiment were of the order of $\Delta\psi$~=~$0.5-1.0^\circ$. It was not possible to precisely determine the angle of rotation~$\psi$ from the step count due to the working principle of the goniometer. However, using the neutron diffractometer during the thermalization of the sample allowed for a quantification of better than $\Delta\psi$~=~0.01$^\circ$ via the vertical tilt of  (0,~$k$,~$l$) nuclear Bragg peaks.

In order to test the functionality of the setup prior to the neutron scattering experiment, we attached a miniature Hall sensor (type \textit{HS-80} from \textit{Advanced Hall Sensors~Ltd.}) to the sample platform. This allowed us to use the magnetic field of the cryomagnet or even the stray field of a permanent magnet, fixed to the outside of the cryostat, to monitor the sample rotation.

\subsection*{\normalsize\normalfont \textit{Supplementary methods}\\\textbf{Preparation of the \textit{Q}-phase state}}

The strong heat load during rotation of the piezoelectric \textit{attocube} sample rotator caused the sample temperature to raise above the superconducting transition temperature, thereby destroying the \textit{Q}-phase state. We exploited this fact to rule out any history-dependent effects that might influence the domain switching shown in figure~3a. First, we moved the goniometer and raised the sample temperature above $T_{\rm s}$~=~1.5~K. We utilized the sample thermometer as a heater, whenever the goniometer did not generate sufficient heating. Then, we realigned the single-crystal and determined the angle of rotation~$\psi$ (see Fig.~3b). With the sample still in the normal state we lowered the magnetic field from $\mu_0 H$~=~11.4~T to 9.0~T. Then, we waited until the superconducting state was entered through the second-order phase transition and a sample temperature of $T_{\rm s}$~=~100~mK was reached. At this point, the sample is in the non-magnetic main \textit{d}-wave superconducting state, but not yet in the \textit{Q}-phase state (see inset of Fig. 1b). Finally, the $Q$-phase was entered through another second-order phase transition at $\mu_0 H_{\rm Q}\approx$~9.8~T by ramping the magnetic field from $\mu_0 H$~=~9.0~T to 11.4~T, while keeping the temperature around  $T_{\rm s}$~=~100~mK.

To measure the hysteresis of the domain switching (see Fig. 3c,d) we had to be very careful not to destroy the \mbox{\textit{Q}-phase} state by heating during movement of the goniometer. The piezo was driven for only about 3~seconds, which corresponds to a rotation angle of about $\Delta\psi$~=~0.04$^\circ$ and caused a temperature increase at the sample of less than 20~mK. In order to assure reproducible measurement conditions, the sample was then allowed to cool to $T_{\rm s}$~=~100~mK, which took about 15~minutes.\\

\end{document}